\documentclass[showpacs,showkeys,twoside,eqsecnum,twocolumn]{revtex4}

\usepackage{amsfonts,amsmath,amssymb,bm,graphicx,dsfont}

\begin{document}

\preprint{E-print archive: hep-ph/0610047}

\title{Evolution of Mixed Particles Interacting with Classical Sources}

\author{Maxim Dvornikov}
\affiliation{Department of Physics, P.O. Box 35, FIN-40014, University of
Jyv\"{a}skyl\"{a}, Finland}
\email{dvmaxim@cc.jyu.fi}
\affiliation{IZMIRAN, 142190, Troitsk, Moscow region, Russia} 
%
\affiliation{Departamento de F\'{i}sica y Centro de Estudios
Subat\'{o}micos, Universidad T\'{e}cnica Federico Santa Mar\'{i}a,
Casilla 110-V, Valpara\'{i}so, Chile}

\date{\today}

\begin{abstract}
We study the systems of scalar and spinor particles with mixing emitted by external 
classical sources. The particles wave functions exactly accounting for external 
sources are obtained directly from the Lorentz invariant wave equations in 
(3+1)-dimensional space-time. Then we discuss sources which are localized in space 
and emit harmonic radiation. We obtain that the considered scalar and spinor fields 
can be converted from one type to another due to the presence of the vacuum mixing. 
This phenomenon is shown to be analogous to neutrino flavor oscillations in vacuum 
since the calculated transition and survival probabilities coincide with the 
corresponding expressions for neutrino oscillations. When we discuss the evolution 
of spinor particles, the situation of general mass matrix, which has both Dirac and 
Majorana mass terms, is studied. In this case we analyze the appearance of 
antiparticles in the initial particles beam. The relevance of the developed 
technique to the description of neutrino flavor oscillations is examined.
\end{abstract}



\maketitle

\section{Introduction}

The studying of the evolution of elementary particles systems with mixing is an 
important problem of high energy physics. Nowadays we have received the 
experimental confirmation that there exists the mixing in the lepton sector of the 
standard model (see, e.g., Ref.~\cite{experiments}) which leads to the transitions 
between different neutrino flavors, i.e. to neutrino flavor oscillations.

Recently considerable efforts were made to theoretically describe the evolution of 
flavor neutrinos with mixing in order to substantiate the results of the quantum 
mechanical approach to the problem in question (see Ref.~\cite{quantmechth,Kob80}). 
The issue of the description of neutrino flavor oscillations with help of field 
theory methods is important since some problems arise when the quantum mechanics is 
used to describe neutrino oscillations. For instance, it is not clear (i) whether 
energies or momenta of different neutrino eigenstates are equal, (ii) whether 
neutrino oscillations occur in space or in time, (iii) whether it is important to 
take into account the multicomponent structure of neutrino wave functions etc. Some 
of the unresolved questions of the neutrino flavor oscillations problem are also 
briefly outlined in Ref.~\cite{Akh06}.

It is necessary to mention that in 
Refs.~\cite{Kob82,GiuKimLeeLee93,GriSto96,BlaVit95,Bla07,LiLiu06,Car00,Dol05,Fie06} 
the quantum field theory was applied for the studying of neutrino flavor 
oscillations in vacuum. In Ref.~\cite{Kob82} the authors studied neutrino 
oscillations with the most general neutrino mass matrix and derived sum rules in 
neutrino oscillations. The authors of Refs.~\cite{GiuKimLeeLee93,GriSto96} 
discussed neutrino oscillations in a similar way as in Ref.~\cite{Kob82}. In those 
works the evolution of neutrino \emph{mass} eigenstates was considered, i.e. one 
formulates neutrino oscillations problem \emph{without} resort to \emph{flavor} 
eigenstates. It should be noticed that, in a sense, the problem of neutrino flavor 
oscillations is incorrectly formulated because in various
processes where a neutrino participates, emitted particles can
belong to several neutrino generations. For example, together with
muon neutrinos in a pion decay, a small number of electron
neutrinos is produced (see also Sec.~\ref{CONCLUSION}). Hence we
should rewrite neutrino emission and detection vertices in
terms of \emph{mass} eigenstates. Then we can examine neutrino emission and 
detection processes by studying incoming and outgoing charged leptons as in 
Refs.~\cite{Kob82,GiuKimLeeLee93,GriSto96}.  These particles are not mixed and thus 
they can serve as properly defined \emph{in} and \emph{out} states.

In Ref.~\cite{BlaVit95} the corrections to the quantum mechanical transition 
probability formula were obtained when the evolution of Fock states of flavor 
neutrinos was examined. In Ref.~\cite{Bla07} it was demonstrated that neutrino 
oscillations can be described in terms of entangled flavor states. Several 
processes involving neutrino weak states, constructed in Ref.~\cite{BlaVit95}, were 
calculated in Ref.~\cite{LiLiu06}. The incompatible results were revealed to appear 
if one uses the inequivalent vacua model proposed in Ref.~\cite{BlaVit95}. The 
$S$-matrix approach was used in Ref.~\cite{Car00} to examine the coherence problem 
in neutrino flavor oscillations. It was shown in Ref.~\cite{Dol05} that neutrino 
oscillations could be suppressed in a gedanken experiment where a neutrino is 
produced in a crystal by a relativistic electron.  Neutrino flavor oscillations 
were analyzed in Ref.~\cite{Fie06} using the path integral method. Finally it 
should be mentioned that various theoretical approaches to neutrino flavor 
oscillations were reviewed in Ref.~\cite{Beu03} which also contains the 
comprehensive bibliography.

We demonstrated in Ref.~\cite{Dvo} that neutrino flavor oscillations in vacuum, in 
matter and in electromagnetic fields of various configurations could be described 
in frames of the relativistic quantum mechanics approach. In those works we formulated the initial condition problem 
and studied the \emph{time} evolution of neutrino \emph{mass} eigenstates using 
Pauli-Jordan function, or its analog taking into account an external field. As a 
result we received the transition probabilities depending on \emph{time}. Therefore 
that method is the most natural generalization, which accounts for the coordinate 
dependence of neutrinos wave functions, of the quantum mechanical treatment of 
neutrino oscillations. The weakness of that approach consists in the fact that to 
obtain a stable oscillations picture one should prepare rather broad, in space, 
initial wave packet of a \emph{flavor} neutrino. Such initial conditions are very 
difficult to implement. In order to overwhelm that difficulty one has to consider 
more realistic models for neutrino emission processes.

This paper continues the series of our publications on the theory of neutrino 
flavor oscillations. The purpose of the present work is to provide a deeper 
understanding of neutrino flavor oscillations in vacuum. We study the evolution of 
mixed scalar (Sec.~\ref{SCALAR}) and spinor (Sec.~\ref{SPINOR}) particles emitted 
by classical sources. We start from Lorentz invariant Lagrangians in 
$(3+1)$-dimensional space-time. Therefore all the results are valid for arbitrary 
energies of emitted particles. The fields distributions of scalar and spinor 
particles exactly accounting for the external sources are obtained. Then for the 
external fields with the specific time and coordinates dependences we derive the 
expressions for energy densities, for scalar fields, and intensities, for spinor 
fields. These quantities are examined in the limit of rapidly oscillating external 
fields. It is shown that the considered expressions coincide with the common 
formulae for transition and survival probabilities of neutrino flavor oscillations 
in vacuum. Moreover in Sec.~\ref{GMT} we study the evolution of fermion fields with 
the general mass matrix and examine the appearance of antiparticles. We discuss our 
results in Sec.~\ref{CONCLUSION} and suggest that the developed formalism is a 
theoretical model for the process of neutrino flavor oscillations. The 
approximations made in deriving of major results are considered in 
Appendix~\ref{ERROR}.

\section{Evolution of scalar particles\label{SCALAR}}

In this section we study the dynamics of $N$ complex scalar particles 
$\bm{\varphi}=(\varphi_1,\dots,\varphi_N)$ with mixing. The scalar particles in 
question are supposed to interact with the external classical fields $f_\lambda$. 
The Lagrangian for this system has the following form:
\begin{align}\label{LagrScal}
  \mathcal{L}(\bm{\varphi})= &
  \sum_{\lambda=1}^{N}
  \partial_\mu\varphi_\lambda^\dag\partial^\mu\varphi_\lambda -
  \sum_{\lambda\lambda'=1}^{N}
  m_{\lambda\lambda'}^2 \varphi_\lambda^\dag \varphi_{\lambda'}
  \notag
  \\
  & +
  \sum_{\lambda=1}^{N}
  (f_\lambda^\dag\varphi_\lambda+\varphi_\lambda^\dag f_\lambda),
\end{align}
where $(m_{\lambda\lambda'}^2)$ is mass martix. We suppose that it is real and 
symmetric. The non-diagonal elements of this matrix are called the vacuum mixing 
terms. The external sources $f_\lambda=f_\lambda(\mathbf{r},t)$ in 
Eq.~\eqref{LagrScal} are taken to be arbitrary functions of coordinates and time.

To describe the evolution of the considered system we make the matrix 
transformation of the Lagrangian~\eqref{LagrScal} and introduce the new set of the 
scalar fields $u_a$,
\begin{equation}\label{scalarmasseigen}
  \varphi_\lambda=
  \sum_{a=1}^{N} U_{\lambda a} u_a,
  \quad
  \varphi_\lambda^\dag=
  \sum_{a=1}^{N}u_a^\dag U_{a \lambda}^\dag,
\end{equation}
in such a way to diagonalize the matrix $(m_{\lambda\lambda'}^2)$, i.e. to 
eliminate the vacuum mixing terms in Eq.~\eqref{LagrScal}. It should be noted that 
the matrix $(U_{\lambda a})$ in Eq.~\eqref{scalarmasseigen} is
orthogonal.

The Lagrangian expressed in terms of the fields $\mathbf{u}=(u_1,\dots,u_N)$ has 
the following form:
\begin{align}
  \mathcal{L}(\mathbf{u})= &
  \sum_{a=1}^{N}
  (\partial_\mu u_a^\dag \partial^\mu u_a - m_a^2 u_a^\dag u_a)
  \notag
  \\
  & +
  \sum_{a=1}^{N}(g_a^\dag u_a+u_a^\dag g_a),
\end{align}
where $m_a$ are the masses of the fields $u_a$. The new sources $g_a$ are obtained 
with help of Eq.~\eqref{scalarmasseigen},
\begin{equation}
  g_a=\sum_{\lambda=1}^{N} U_{a \lambda}^\dag f_\lambda,
  \quad
  g_a^\dag=\sum_{\lambda=1}^{N} f_\lambda^\dag U_{\lambda a}.
\end{equation}
The dynamic equation for the field $u_a$ is the inhomogeneous Klein-Gordon 
equation,
\begin{equation}\label{scalarinhomg}
  (\Box-m_a^2)u_a=-g_a,
\end{equation}
where $\Box=-\partial_\mu \partial^\mu$ is the d'Alembertian operator. It is worth 
mentioning that the squares of masses $m_a^2$ of the fields $u_a$ are the 
eigenvalues of the matrix $(m_{\lambda\lambda'}^2)$.

The solution to Eq.~\eqref{scalarinhomg} for the arbitrary function $g_a$ has the 
form (see, e.g., Ref.~\cite{BogShi60p136})
\begin{equation}\label{masseigensol1}
  u_a(\mathbf{r},t)=\int\mathrm{d}^3\mathbf{r}'\mathrm{d}t'
  D^\mathrm{ret}_a(\mathbf{r}-\mathbf{r}',t-t')g_a(\mathbf{r}',t'),
\end{equation}
where $D^\mathrm{ret}_a(\mathbf{r},t)$ is the retarded Green function. This 
function obeys the equation,
\begin{align}\label{propD}
  & (\Box-m_a^2)D^\mathrm{ret}_a(x)=
  -\delta^4(x),
  \notag
  \\
  &
  D^\mathrm{ret}_a(x)=0
  \quad
  \text{for}
  \quad
  t<0,
\end{align}
where $x^\mu=(t,\mathbf{r})$. The explicit form of the function 
$D^\mathrm{ret}_a(x)$ can be found in Ref.~\cite{BogShi60p651},
\begin{align}\label{retGfunScal}
  D^\mathrm{ret}_a(\mathbf{r},t) & =
  \int\frac{\mathrm{d}^4 p}{(2\pi)^4}
  \frac{e^{\mathrm{i}px}}{m_a^2-p^2+\mathrm{i}\epsilon p^0}
  \notag
  \\
  & =
  \frac{1}{2\pi}\theta(t)
  \left\{
    \delta(s^2)-\theta(s^2)\frac{m_a}{2s}J_1(m_a s)
  \right\},
\end{align}
where $\theta(t)$ is the Heaviside step function, $J_1(z)$ is the first order 
Bessel function, $s=\sqrt{t^2-r^2}$ and $r=|\mathbf{r}|$.

Now it is necessary to specify the time and spatial coordinates dependence of the 
external fields $f_\lambda(\mathbf{r},t)$. We assume that the sources are localized 
in space and emit harmonic radiation,
\begin{equation}
  f_\lambda(\mathbf{r},t)=
  \theta(t)f_\lambda^{(0)} e^{-\mathrm{i}E t}\delta^3(\mathbf{r}),
\end{equation}
where $f_\lambda^{(0)}$ is the amplitude of the source and $E$ is its frequency, 
the analog of the energy of emitted particles. In this case 
Eq.~\eqref{masseigensol1} is rewritten in the following way:
\begin{equation}\label{masseigensol2}
  u_a(\mathbf{r},t)=g_a^{(0)}e^{-\mathrm{i}E t}
  \int_0^t\mathrm{d}\tau
  D^\mathrm{ret}_a(\mathbf{r},\tau)e^{\mathrm{i}E\tau},
\end{equation}
where
\begin{equation*}
  g_a^{(0)}=\sum_{\lambda=1}^{N} U_{a\lambda}^\dag f_\lambda^{(0)},
\end{equation*}
is the time independent component of the function $g_a$.

One can see that two major terms appear in Eq.~\eqref{masseigensol2} [see also 
Eq.~\eqref{retGfunScal}] while integrating over $\tau$
\begin{equation}
  \int_0^t\mathrm{d}\tau
  D^\mathrm{ret}_a(\mathbf{r},\tau)e^{\mathrm{i}E\tau}=
  \mathcal{I}_1+\mathcal{I}_2.
\end{equation}
They are
\begin{align}
  \label{int1}
  \mathcal{I}_1 & =\frac{1}{2\pi}\int_0^t\mathrm{d}\tau
  \delta(s^2)e^{\mathrm{i}E\tau}=
  \frac{\theta(t-r)}{4\pi r}e^{\mathrm{i}E r},
  \\
  \label{int2}
  \mathcal{I}_2 & =-\frac{m_a}{4\pi}\int_0^t\mathrm{d}\tau
  \theta(s^2)\frac{J_1(m_a s)}{s}e^{\mathrm{i}E\tau}
  \notag
  \\
  & =
  -\frac{m_a}{4\pi}\theta(t-r)\int_0^{x_m}\mathrm{d}x
  \frac{J_1(m_a x)}{\sqrt{r^2+x^2}}e^{\mathrm{i}E\sqrt{r^2+x^2}},
\end{align}
where $x_m=\sqrt{t^2-r^2}$ and $s=\sqrt{\tau^2-r^2}$. It is interesting to notice 
that both $\mathcal{I}_1$ and $\mathcal{I}_2$ in Eqs.~\eqref{int1} and~\eqref{int2} 
are equal to zero if $t<r$. It means that the initial perturbation from a source 
placed at the point $\mathbf{r}=0$ reaches a detector placed at the point 
$\mathbf{r}$ only after the time $t>r$.

Despite the integral $\mathcal{I}_1$ in Eq.~\eqref{int1} is expressed in terms of 
the elementary functions, the integral $\mathcal{I}_2$ in Eq.~\eqref{int2} cannot 
be computed analytically for arbitrary $r$ and $t$. However some reasonable 
assumptions can be made to simplify the considered expression. An observer is at 
the fixed distance from a source. As we have already mentioned one detects a signal 
starting from $t>r$. It is obvious that a non-stationary rapidly oscillating signal 
is detected when the wave front just arrives to a detector, i.e. when $t \gtrsim 
r$. The situation is analogous to waves propagating on the water surface. 
Therefore, if we suppose that one starts observing particles when the 
non-stationary signal attenuates, i.e. $t\gg r$ or $x_m\to\infty$, we can avoid 
relaxation phenomena. In this case the integral $\mathcal{I}_2$ can be computed 
analytically,
\begin{equation}\label{int2fin}
  \mathcal{I}_2=-\frac{1}{4\pi r}(e^{\mathrm{i}E r}-e^{\mathrm{i}p_a r}),
\end{equation}
where $p_a=\sqrt{E^2-m_a^2}$ is the analog of the particle momentum.

To obtain Eq.~\eqref{int2fin} we use the known values of the integrals,
\begin{align*}
  \int_0^\infty \mathrm{d}x &
  \frac{J_\nu(m x)}{\sqrt{r^2+x^2}}\sin(E\sqrt{r^2+x^2})
  \\
  = &
  \frac{\pi}{2}
  J_{\nu/2}
  \left[
    \frac{r}{2}
    \left(
      E-\sqrt{E^2-m^2}
    \right)
  \right]
  \\
  & \times
  J_{-\nu/2}
  \left[
    \frac{r}{2}
    \left(
      E+\sqrt{E^2-m^2}
    \right)
  \right],
  \\
  \int_0^\infty \mathrm{d}x &
  \frac{J_\nu(m x)}{\sqrt{r^2+x^2}}\cos(E\sqrt{r^2+x^2})
  \\
  = &
  -\frac{\pi}{2}
  J_{\nu/2}
  \left[
    \frac{r}{2}
    \left(
      E-\sqrt{E^2-m^2}
    \right)
  \right]
  \\
  & \times
  N_{-\nu/2}
  \left[
    \frac{r}{2}
    \left(
      E+\sqrt{E^2-m^2}
    \right)
  \right],
\end{align*}
and the fact that the Bessel and Neumann functions of $\pm 1/2$ order
\begin{align*}
  J_{1/2}(z)= N_{-1/2}(z)= & \sqrt{\frac{2}{\pi z}}\sin z,
  \\
  J_{-1/2}(z)= & \sqrt{\frac{2}{\pi z}}\cos z,
\end{align*}
are expressed in terms of the elementary functions. The approximations made in 
derivation of Eq.~\eqref{int2fin} are analysed in Appendix~\ref{ERROR}.

Using Eqs.~\eqref{int1} and~\eqref{int2fin} we obtain the field distribution of 
$u_a$,
\begin{equation}\label{scalarmassevol}
  u_a(\mathbf{r},t)=e^{-\mathrm{i}E t+\mathrm{i}p_a r}
  \frac{g_a^{(0)}}{4\pi r}.
\end{equation}
Then it is necessary to describe the evolution of the fields $\varphi_\lambda$. 
With help of Eqs.~\eqref{scalarmasseigen} and~\eqref{scalarmassevol} one receives 
the field distribution of $\varphi_\lambda$ in the following form:
\begin{equation}\label{scalarflavevol}
  \varphi_\lambda(\mathbf{r},t)=
  e^{-\mathrm{i}E t}
  \sum_{a\lambda'=1}^{N}
  U_{\lambda a}U_{a\lambda'}^\dag
  e^{\mathrm{i}p_a r}\frac{f_{\lambda'}^{(0)}}{4\pi r}.
\end{equation}
We remind that this formula is valid if $t\gg r$.

To be more elucidative we consider the system of only two scalar fields, 
$\varphi_1$ and $\varphi_2$. In this case the matrix $(U_{\lambda a})$ in 
Eq.~\eqref{scalarmasseigen} is parametrized with help of the mixing angle $\theta$,
\begin{equation}\label{Mtheta}
  (U_{\lambda a})=
  \begin{pmatrix}
    \cos\theta & -\sin\theta \\
    \sin\theta & \cos\theta
  \end{pmatrix}.
\end{equation}
We also adopt the following amplitudes of the external sources: $f_1^{(0)}=0$ and 
$f_2^{(0)}\equiv f\neq 0$. If we discussed the evolution of flavor neutrinos, that 
is in fact done in Sec.~\ref{SPINOR}, such a choice of the external fields would 
mean that $\varphi_1$ would be a muon or a $\tau$-neutrino and $\varphi_2$ -- an 
electron neutrino. There would be a source of only electron neutrinos at 
$\mathbf{r}=0$ and a detector at the distance $r$ from a source.

Using Eqs.~\eqref{scalarflavevol} and~\eqref{Mtheta} we obtain the fields 
distributions of $\varphi_{1,2}$ in the following form:
\begin{align}\label{phi12evol}
  \varphi_1(\mathbf{r},t)= &
  \sin\theta \cos\theta
  e^{-\mathrm{i}E t}
  \frac{f}{4\pi r}
  (e^{\mathrm{i}p_1 r}-e^{\mathrm{i}p_2 r}),
  \notag
  \\
  \varphi_2(\mathbf{r},t)= &
  e^{-\mathrm{i}E t}
  \frac{f}{4\pi r}
  (\sin^2\theta e^{\mathrm{i}p_1 r}+\cos^2\theta e^{\mathrm{i}p_2 r}).
\end{align}
It was demonstrated in Ref.~\cite{Dvo} that the measurable quantity for a scalar 
field is the energy density rather than the field distribution. Constructing the 
time component $T^{00}$ of the energy momentum tensor of $\varphi_\lambda$ we can 
write down the energy density in the following way:
\begin{align}
  \mathcal{H}_\lambda(\mathbf{r},t) & = T^{00}[\varphi_\lambda(\mathbf{r},t)]
  \notag
  \\
  & =
  |\dot{\varphi}_\lambda|^2+|\bm{\nabla}\varphi_\lambda|^2+
  m_{\lambda\lambda}^2|\varphi_\lambda|^2.
\end{align}
Now let us discuss the high frequency approximation of the obtained results: $E\gg 
m_{1,2}$. If we again referred to the flavor neutrinos evolution (see 
Sec.~\ref{SPINOR}), it would mean that ultrarelativistic neutrinos would be 
emitted. Decomposing the "momenta" $p_a\approx E-m_a^2/(2E)$ in 
Eqs.~\eqref{phi12evol} and taking into account that $\mathcal{H}_\lambda\approx 
|\dot{\varphi}_\lambda|^2+|\bm{\nabla}\varphi_\lambda|^2\approx 
2E^2|\varphi_\lambda|^2$ one receives the expressions for the energy densities
\begin{align}\label{phi12endens}
  \mathcal{H}_1(r)= & 2E^2
  \frac{|f|^2}{(4\pi r)^2}
  \\
  & \times
  \left\{
    \sin^2(2\theta)\sin^2
    \left(
      \frac{\Delta m^2}{4E}r
    \right)+
    \mathcal{O}
    \left(
      \frac{m_a^2}{E^2}
    \right)
  \right\},
  \notag
  \\
  \mathcal{H}_2(r)= & 2E^2
  \frac{|f|^2}{(4\pi r)^2}
  \notag
  \\
  & \times
  \left\{
    1-\sin^2(2\theta)\sin^2
    \left(
      \frac{\Delta m^2}{4E}r
    \right)+
    \mathcal{O}
    \left(
      \frac{m_a^2}{E^2}
    \right)
  \right\},
  \notag
\end{align}
where we use the common notation $\Delta m^2=m_1^2-m_2^2$.

With help of Eqs.~\eqref{phi12endens} we obtain the transition and survival 
probabilities in the following way:
\begin{align}
  \label{tranprscal}
  P_{2\to 1}^{(\mathrm{scalar})}(r) \sim &
  \sin^2(2\theta)\sin^2
  \left(
    \frac{\Delta m^2}{4E}r
  \right),
  \\
  \label{survprscal}
  P_{2\to 2}^{(\mathrm{scalar})}(r) \sim &
  1-\sin^2(2\theta)\sin^2
  \left(
    \frac{\Delta m^2}{4E}r
  \right).
\end{align}
It can be seen that Eqs.~\eqref{tranprscal} and~\eqref{survprscal} coincide with 
the usual formulae for transition and survival probabilities of neutrino flavor 
oscillations in vacuum that were obtained in frames of the quantum mechanical 
approach~\cite{quantmechth,Kob80}.

\section{Evolution of spinor particles\label{SPINOR}}

In this section we consider the problem analogous to that we discuss in 
Sec.~\ref{SCALAR}, namely the evolution of $N$ spinor particles 
$\bm{\nu}=(\nu_1,\dots,\nu_N)$ with mixing under the influence of the external 
classical fields $f_\lambda^\mu$.

The Lorentz invariant vacuum, i.e. without external fields, Lagrangian generally 
has the form,
\begin{align}\label{Lagrvac}
  \mathcal{L}(\bm{\nu})= &
  \sum_{\lambda=1}^{N}
  \left(
    \overline{\nu_\lambda^\mathrm{L}}
    \mathrm{i}\gamma^\mu\partial_\mu
    \nu_\lambda^\mathrm{L}+
    \overline{\nu_\lambda^\mathrm{R}}
    \mathrm{i}\gamma^\mu\partial_\mu
    \nu_\lambda^\mathrm{R}
  \right)
  \notag
  \\
  & -
  \sum_{\lambda\lambda'=1}^{N}
  \Big(
    m_{\lambda\lambda'}^\mathrm{D}
    \overline{\nu_\lambda^\mathrm{L}}\nu_{\lambda'}^\mathrm{R}+
    m_{\lambda\lambda'}^\mathrm{L}
    \left(
      \nu_\lambda^\mathrm{L}
    \right)^\mathrm{T}
    C \nu_{\lambda'}^\mathrm{L}
    \notag
    \\
    & +
    m_{\lambda\lambda'}^\mathrm{R}
    \left(
      \nu_\lambda^\mathrm{R}
    \right)^\mathrm{T}
    C \nu_{\lambda'}^\mathrm{R}
    +\text{h.c.}
  \Big),
\end{align}
where $\nu_\lambda^\mathrm{L,R}=(1/2)(1 \mp \gamma^5)\nu_\lambda$ are the chiral 
projections, $\left( m_{\lambda\lambda'}^\mathrm{D} \right)$, $\left( 
m_{\lambda\lambda'}^\mathrm{L} \right)$, $\left( m_{\lambda\lambda'}^\mathrm{R} 
\right)$ are the Dirac as well as (left and right) Majorana mass matrices and 
$C=\mathrm{i}\gamma^2\gamma^0$ is the charge conjugation matrix. These mass 
matrices should satisfy certain requirements which are discussed in 
Ref.~\cite{Kob80}. The Lagrangian~\eqref{Lagrvac} is CPT-invariant. In this section 
we adopt the notations for Dirac matrices as in Ref.~\cite{ItzZub80}. The analogous 
mass terms are generated in theoretical models based on the type-II seesaw 
mechanism (see, e.g., Ref.~\cite{SchVal80}).

We supply the Lagrangian~\eqref{Lagrvac} with the term describing particles 
interaction with external sources. Suppose that a particle is involved in the 
left-handed currents interactions,
\begin{equation}\label{Lagrint}
  \mathcal{L}_\mathrm{int}=
  \sum_{\lambda=1}^{N}
  (\bar{\ell}_\lambda\gamma_\mu^\mathrm{L}\nu_\lambda f_\lambda^{\dag\mu}+
  \bar{\nu}_\lambda\gamma_\mu^\mathrm{L}\ell_\lambda f_\lambda^\mu),
\end{equation}
where $\gamma_\mu^\mathrm{L}=\gamma_\mu(1-\gamma^5)/2$. The external fields 
$f_\lambda^\mu=f_\lambda^\mu(\mathbf{r},t)$ are supposed to be arbitrary functions. 
The physical analog of the Lagrangian~\eqref{Lagrint} is the system of flavor 
neutrinos $\nu_\lambda$ interacting with matter by means of the electroweak 
interactions. In this case $\ell_\lambda$ is the charged $\mathrm{SU}(2)$ 
isodoublet partner of $\nu_\lambda$. If we study the neutrino emission in a process 
like the inverse $\beta$-decay: $p+e^{-}\to\nu_e+n$, the external fields in 
Eq.~\eqref{Lagrint} are (see, e.g., Ref.~\cite{Oku90})
\begin{align}\label{fmuexample}
  f_{\nu_e}^\mu = & -\sqrt{2}G_F
  \bar{\varPsi}_n \gamma^\mu (1-\alpha\gamma^5) \varPsi_p,
  \notag
  \\
  f_{\nu_\mu}^\mu = & 0,
  \quad
  f_{\nu_\tau}^\mu = 0,
\end{align}
where $\varPsi_p$ and $\varPsi_n$ are the wave functions of a proton and a neutron, 
$\alpha\approx 1.25$ and $G_\mathrm{F}$ is the Fermi constant. In 
Eq.~\eqref{fmuexample} we assume that a source consists of electrons, protons and 
neutrons.

After the diagonalization of the whole mass term of the Lagrangian~\eqref{Lagrvac} 
one obtains $2N$ Majorana particles with different masses~\cite{Kob80}. In the 
following we discuss the case when only Dirac mass term is presented. Then we study 
the general situation.

\subsection{Dirac mass term\label{DMT}}

In this section we suppose that only Dirac mass matrix is non-zero in 
Eq.~\eqref{Lagrvac}. Analogously to the discussion in Sec.~\ref{SCALAR} we 
introduce the new set of spinor fields $\bm{\psi}=(\psi_1,\dots,\psi_N)$ by means 
of the matrix transformation. In our situation the mixing matrix is the $N \times 
N$ unitary matrix $(U_{\lambda a})$,
\begin{equation}\label{spinnupsirel}
  \nu_\lambda=\sum_{a=1}^N
  U_{\lambda a} \psi_a,
  \quad
  \bar{\nu}_\lambda=\sum_{a=1}^N
  \bar{\psi}_a U_{a \lambda}^\dag.
\end{equation}
The corresponding mass eigenstates $\psi_a$ are Dirac particles.

When we transform the sum of the Lagrangians~\eqref{Lagrvac} and~\eqref{Lagrint} 
using Eq.~\eqref{spinnupsirel}, it is rewritten in the following way:
\begin{align}\label{spinmassLarg}
  \mathcal{L}(\bm{\psi})= &
  \sum_{a=1}^{N}
  \bar{\psi}_a (\mathrm{i}\gamma^\mu\partial_\mu - m_a) \psi_a
  \notag
  \\
  & +
  \sum_{a=1}^{N}(\bar{\Psi}_a\psi_a+\bar{\psi}_a\Psi_a),
\end{align}
where $\Psi_a$ is the external source for the fermion $\psi_a$,
\begin{equation}\label{spinmasssour}
  \Psi_a=
  \sum_{\lambda=1}^{N}
  U_{a \lambda}^\dag
  \gamma_\mu^\mathrm{L} \ell_\lambda f_\lambda^\mu,
  \quad
  \bar{\Psi}_a=
  \sum_{\lambda=1}^{N}
  \bar{\ell}_\lambda \gamma_\mu^\mathrm{L} f_\lambda^{\dag\mu}
  U_{\lambda a}.
\end{equation}
Using Eqs.~\eqref{spinmassLarg} and~\eqref{spinmasssour} we receive the 
inhomogeneous Dirac equation for the fermion $\psi_a$,
\begin{equation}\label{inhomDir}
  (\mathrm{i}\gamma^\mu\partial_\mu-m_a)\psi_a=-\Psi_a.
\end{equation}
We mention that the masses $m_a$ are the eigenvalues of the matrix 
$(m_{\lambda\lambda'}^\mathrm{D})$.

The solution to Eq.~\eqref{spinmasssour} for the arbitrary spinor $\Psi_a$ is 
expressed with help of the retarded Green function for a spinor field (see again 
Ref.~\cite{BogShi60p136}),
\begin{equation}\label{spinmasseigensol1}
  \psi_a(\mathbf{r},t)=\int\mathrm{d}^3\mathbf{r}'\mathrm{d}t'
  S^\mathrm{ret}_a(\mathbf{r}-\mathbf{r}',t-t')\Psi_a(\mathbf{r}',t').
\end{equation}
The explicit form of $S^\mathrm{ret}_a(\mathbf{r},t)$ can be also found in 
Ref.~\cite{BogShi60p136},
\begin{equation}\label{spinretGfun}
  S^\mathrm{ret}_a(\mathbf{r},t)=
  (\mathrm{i}\gamma^\mu\partial_\mu+m_a)D^\mathrm{ret}_a(\mathbf{r},t),
\end{equation}
where the function $D^\mathrm{ret}_a(\mathbf{r},t)$ is given in Sec.~\ref{SCALAR}.

To proceed in further calculations it is necessary to define the behavior of the 
external sources. Let us suppose that the external fields depend on time and 
spatial coordinates in the similar way as in Sec.~\ref{SCALAR}, namely
\begin{equation}\label{spinflavsour}
  \ell_\lambda(\mathbf{r},t) f_\lambda^\mu(\mathbf{r},t)=
  \theta(t)l_\lambda f_\lambda^{(0)\mu}
  e^{-\mathrm{i}E t}\delta^3(\mathbf{r}),
\end{equation}
where $l_\lambda=\ell^{(0)}_\lambda$ is the time independent component of the 
spinor $\ell_\lambda$ and $f_\lambda^{(0)\mu}$ is the amplitude of the function 
$f_\lambda^\mu(\mathbf{r},t)$. Using Eqs.~\eqref{spinmasssour} 
and~\eqref{spinflavsour} we obtain for $\Psi_a$,
\begin{align}\label{spinamplchi}
  \Psi_a(\mathbf{r},t)= & \theta(t)\Psi_a^{(0)}
  e^{-\mathrm{i}E t}\delta^3(\mathbf{r}),
  \notag
  \\
  \Psi_a^{(0)}= &
  \sum_{\lambda=1}^{N} U_{a \lambda}^\dag \gamma_\mu^\mathrm{L}
  l_\lambda f_\lambda^{(0)\mu}.
\end{align}
With help of Eqs.~\eqref{spinretGfun}-\eqref{spinamplchi} we can rewrite    
Eq.~\eqref{spinmasseigensol1} in the following way:
\begin{align}
  \psi_a(\mathbf{r},t)= &
  \left(
    \mathrm{i}\gamma^\mu\frac{\partial}{\partial x^\mu}+m_a
  \right)
  \notag
  \\
  & \times
  \left\{
    e^{-\mathrm{i}E t}
    \int_0^t\mathrm{d}\tau
    D^\mathrm{ret}_a(\mathbf{r},\tau)e^{\mathrm{i}E\tau}
  \right\}
  \Psi_a^{(0)}.
\end{align}
Now it is possible to use the technique developed in Sec.~\ref{SCALAR}. Finally we 
obtain the field distribution of the fermion $\psi_a$,
\begin{equation}\label{spinmasseigensol2}
  \psi_a(\mathbf{r},t)=
  e^{-\mathrm{i}E t+\mathrm{i}p_a r} O_a
  \frac{\Psi_a^{(0)}}{4\pi r},
\end{equation}
where $O_a=\gamma^0E-(\bm{\gamma}\mathbf{n})p_a+m_a$ and $\mathbf{n}$ is the unit 
vector towards a detector. It should be noted that in deriving 
Eq.~\eqref{spinmasseigensol2} we differentiate only exponential rather than the 
factor $1/r$ because the derivative of $1/r$ is proportional to $1/r^2$. Such a 
term is negligible at large distances from a source. We also remind that 
Eq.~\eqref{spinmasseigensol2} is valid for $t\gg r$.

Let us turn to the description of the evolution of the fields $\nu_\lambda$. Using 
Eqs.~\eqref{spinnupsirel} and~\eqref{spinmasseigensol2}, we obtain the 
corresponding field distribution,
\begin{equation}\label{spinflavfin}
  \nu_\lambda(\mathbf{r},t)=
  e^{-\mathrm{i}E t}
  \sum_{a\lambda'=1}^{N}
  U_{\lambda a} U_{a\lambda'}^\dag
  e^{\mathrm{i}p_a r}
  O_a\gamma_\mu^\mathrm{L} l_{\lambda'}
  \frac{f_{\lambda'}^{(0)\mu}}{4\pi r}.
\end{equation}

Analogously to Sec.~\ref{SCALAR} we study the evolution of only two fermions. The 
mixing matrix is given in Eq.~\eqref{Mtheta}. We also choose the amplitudes of the 
sources in the following way: $f_1^{(0)\mu}=0$, $f_2^{(0)\mu}\equiv f^\mu \neq 0$ 
and $l_2\equiv l$. The reason for such a choice is similar to that in 
Sec.~\ref{SCALAR}. If we suppose that $\nu_1$ corresponds to a muon or a 
$\tau$-neutrino and $\nu_2$ -- to an electron neutrino, then it means that there is 
a source of only electron neutrinos and we observe particles at the distance $r$ 
from a source. Using Eqs.~\eqref{Mtheta} and~\eqref{spinflavfin} we obtain the 
fields distributions of each of the particles $\nu_{1,2}$,
\begin{align}\label{spinnu12wf}
  \nu_1(\mathbf{r},t)= &
  \sin\theta\cos\theta
  e^{-\mathrm{i}E t}
  \frac{f^\mu}{4\pi r}
  \notag
  \\
  & \times
  (e^{\mathrm{i}p_1 r}O_1-e^{\mathrm{i}p_2 r}O_2)
  \gamma_\mu^\mathrm{L} l,
  \notag
  \\
  \nu_2(\mathbf{r},t)= &
  e^{-\mathrm{i}E t}
  \frac{f^\mu}{4\pi r}
  \notag
  \\
  & \times
  (\sin^2\theta e^{\mathrm{i}p_1 r}O_1+\cos^2\theta e^{\mathrm{i}p_2 r}O_2)
  \gamma_\mu^\mathrm{L} l.
\end{align}
In the following we discuss the high frequency approximation: $E\gg m_{1,2}$. It 
again corresponds to the emission of ultrarelativistic neutrinos.

It was shown in our previous works~\cite{Dvo} that the measurable quantity for a 
spinor particle is the intensity which is defined by the following expression: 
$I_\lambda(\mathbf{r},t)=|\nu_\lambda(\mathbf{r},t)|^2$. With help of 
Eqs.~\eqref{spinnu12wf} we get the intensities in the form
\begin{align}\label{spinint1}
  I_1(r)= & -2E^2
  \frac{f^{\dag\mu} f^\nu}{(4\pi r)^2}
  \langle T_{\mu\nu} \rangle
  \\
  & \times
  \left\{
  \sin^2(2\theta)\sin^2
    \left(
      \frac{\Delta m^2}{4E}r
    \right)+
    \mathcal{O}
    \left(
      \frac{m_a^2}{E^2}
    \right)
  \right\},
  \notag
  \\
  I_2(r)= & -2E^2
  \frac{f^{\dag\mu} f^\nu}{(4\pi r)^2}
  \langle T_{\mu\nu} \rangle
  \notag
  \\
  & \times
  \notag
  \left\{
    1-\sin^2(2\theta)\sin^2
    \left(
      \frac{\Delta m^2}{4E}r
    \right)+
    \mathcal{O}
    \left(
      \frac{m_a^2}{E^2}
    \right)
  \right\},
\end{align}
where $\langle T_{\mu\nu} \rangle= 
l^\dag\gamma_\mu^\dag(\bm{\alpha}\mathbf{n})\gamma_\nu l$ and 
$\bm{\alpha}=\gamma^0\bm{\gamma}$. It is possible to calculate the components of 
the tensor $\langle T_{\mu\nu} \rangle$. They depend on the properties of the 
fermion $\ell_2\equiv\ell$,
\begin{gather}
  \langle T_{00} \rangle=-(\mathbf{v}\mathbf{n}),
  \quad
  \langle T_{0i} \rangle=\langle T_{i0} \rangle^{*}=n_i-
  \mathrm{i}[\mathbf{n}\times\bm{\zeta}]_i,
  \notag
  \\
  \label{Tmunucomp}
  \langle T_{ij} \rangle =
  \delta_{ij}(\mathbf{v}\mathbf{n})-
  (v_i n_j + v_j n_i) +
  \mathrm{i}\varepsilon_{ijk}n_k,
\end{gather}
where $\mathbf{v}=\langle \bm{\alpha} \rangle$ is the velocity of fermion $\ell$, 
$\bm{\zeta}=\langle \bm{\Sigma} \rangle$ is its spin and 
$\bm{\Sigma}=\gamma^5\bm{\alpha}$. It should be noted that, if we study the 
evolution of flavor neutrinos and $\nu_2$ corresponds to an electron neutrino, the 
fermion $\ell$ is an electron.

Let us discuss the simplified case when the spatial components of the four-vector 
$f^\mu$ are equal to zero: $\mathbf{f}=0$. It corresponds to the neutrino emission 
by a non-moving and unpolarized source. Using Eq.~\eqref{Tmunucomp} we can rewrite 
Eq.~\eqref{spinint1} in the following way:
\begin{align}\label{spinint2}
  I_1(r)\approx & 2E^2(\mathbf{v}\mathbf{n})
  \frac{|f^0|^2}{(4\pi r)^2}
  \sin^2(2\theta)\sin^2
  \left(
    \frac{\Delta m^2}{4E}r
  \right),
  \notag
  \\
  I_2(r)\approx & 2E^2(\mathbf{v}\mathbf{n})
  \frac{|f^0|^2}{(4\pi r)^2}
  \notag
  \\
  & \times
  \left\{
    1-\sin^2(2\theta)\sin^2
    \left(
      \frac{\Delta m^2}{4E}r
    \right)
  \right\}.
\end{align}
Now we are able to introduce the transition and survival probabilities on the basis 
of Eq.~\eqref{spinint2},
\begin{align}
  \label{tranprspin}
  P_{2\to 1}^{(\mathrm{spinor})}(r) \sim &
  \sin^2(2\theta)\sin^2
  \left(
    \frac{\Delta m^2}{4E}r
  \right),
  \\
  \label{survprspin}
  P_{2\to 2}^{(\mathrm{spinor})}(r) \sim &
  1-\sin^2(2\theta)\sin^2
  \left(
    \frac{\Delta m^2}{4E}r
  \right).
\end{align}
It should be noted that Eqs.~\eqref{tranprspin} and~\eqref{survprspin} are the same 
as the common formulae for the description of neutrino flavor oscillations in 
vacuum~\cite{quantmechth,Kob80}. These expressions also coincide with 
Eqs.~\eqref{tranprscal} and~\eqref{survprscal} derived for the case of the mixed 
scalar fields.

\subsection{General mass term\label{GMT}}

To begin analyzing the dynamics of mixed fermions with the general mass 
term~\eqref{Lagrvac} we rewrite the interaction Lagrangian~\eqref{Lagrint} in the 
equivalent form,
\begin{equation}\label{LagrintM}
  \mathcal{L}_\mathrm{int}=
  \sum_{\lambda=1}^{N}
  \left(
    \overline{\ell_\lambda^\mathrm{L}} \gamma_\mu \nu_\lambda^\mathrm{L}
    f_\lambda^{\dag\mu}+
    \overline{\nu_\lambda^\mathrm{L}} \gamma_\mu \ell_\lambda^\mathrm{L}
    f_\lambda^\mu
  \right).
\end{equation}
Now we can introduce mass eigenstates $\psi_a$. The analog of 
Eq.~\eqref{spinnupsirel} is (see also Ref.~\cite{Kob80})
\begin{equation}\label{spinnupsirelM}
  \nu_\lambda^\mathrm{L} = \sum_{a=1}^{2N}
  U_{\lambda a} \psi_a^\mathrm{L},
  \quad
  \overline{\nu_\lambda^\mathrm{L}}=\sum_{a=1}^{2N}
  \overline{\psi_a^\mathrm{L}} U_{a \lambda}^\dag.
\end{equation}
where the matrix $(U_{\lambda a})$ is the rectangular $N \times 2N$ matrix. Note 
that it is not unitary. We also mention that the mass eigenstates $\psi_a$ are 
Majorana particles. To implement the complete diagonalization of the mass 
term~\eqref{Lagrvac} it is necessary to use $2N \times 2N$ unitary matrix, i.e. 
along with Eq.~\eqref{spinnupsirelM} one should consider also the transformation of 
the right-handed chiral components of the particles $\nu_\lambda$,
\begin{equation}\label{matrV}
  \nu_\lambda^\mathrm{R} = \sum_{a=1}^{2N}
  V_{\lambda a} \psi_a^\mathrm{R}.
\end{equation}
Note that the matrices $(U_{\lambda a})$ and $(V_{\lambda a})$ in 
Eqs.~\eqref{spinnupsirelM} and~\eqref{matrV} are independent. The modern 
parameterization for these matrices is given in Ref.~\cite{Xin07}.

The Lagrangian for the particles $\psi_a$ takes the form,
\begin{align}\label{spinmassLargM}
  \mathcal{L}(\bm{\psi})= &
  \sum_{a=1}^{2N}
  \overline{\psi_a^\mathrm{L}}
  \mathrm{i}\gamma^\mu\partial_\mu \psi_a^\mathrm{L}
  \notag
  \\
  & -
  \sum_{a=1}^{2N}
  \left(
    m_a \overline{\psi_a^\mathrm{L}} \psi_a^\mathrm{R} -
    \overline{\Psi_a^\mathrm{R}} \psi_a^\mathrm{L} +
    \text{h.c.}
  \right),
\end{align}
where the external sources $\Psi_a \equiv \Psi_a^\mathrm{R}$ have the same form as 
in Eq.~\eqref{spinmasssour}.

Note that Majorana spinors are equivalent to two-component Weil 
spinors~\cite{Cas57}. Hence we can rewrite the spinors $\psi_a^\mathrm{L,R}$ and 
$\Psi_a$ as
\begin{equation}
  \psi_a^\mathrm{L} =
  \begin{pmatrix}
    0 \\
    \eta_a
  \end{pmatrix},
  \quad
  \psi_a^\mathrm{R} =
  \begin{pmatrix}
    \mathrm{i}\sigma_2\eta_a^{*{}} \\
    0
  \end{pmatrix},
  \quad
  \Psi_a =
  \begin{pmatrix}
    \phi_a \\
    0
  \end{pmatrix},
\end{equation}
where
\begin{equation}\label{phisour}
  \phi_a=
  \sum_{\lambda=1}^{N}
  U_{a \lambda}^\dag
  \chi_\lambda f_\lambda^0.
\end{equation}
To obtain Eq.~\eqref{phisour} we suppose as in Sec.~\ref{DMT} that the vectors 
$f_\lambda^\mu$ have only time component $f_\lambda^0$. We also assume that $\left( 
\ell^\mathrm{L}_\lambda \right)^\mathrm{T}=(0,\chi_\lambda)$. To derive 
Eq.~\eqref{phisour} it is crucial that only left-handed currents interactions are 
presented in Eqs.~\eqref{Lagrint} and~\eqref{spinmasssour}.

It is useful to rewrite the Lagrangian~\eqref{spinmassLargM} in terms of the  
two-component spinors $\bm{\eta} = (\eta_1, \dots, \eta_{2N})$, $\bm{\eta}^{*{}} = 
(\eta_1^{*{}}, \dots, \eta_{2N}^{*{}})$ and $(\phi_1, \dots, 
\phi_{2N})$~\cite{FukYan03},
\begin{align}\label{Largetaeta*}
  \mathcal{L}(\bm{\eta},\bm{\eta}^{*{}}) = &
  \sum_{a=1}^{2N}
  \mathrm{i} \eta_a^\dag (\partial_t - \bm{\sigma}\bm{\nabla}) \eta_a
  \notag
  \\
  & +
  \sum_{a=1}^{2N}
  \left(
    \frac{\mathrm{i}}{2} m_a \eta_a^\dag \sigma_2 \eta_a^{*{}} +
    \eta_a^\dag \phi_a + \text{h.c.}
  \right).
\end{align}
Using Eq.~\eqref{Largetaeta*} we can receive the wave equation for two-component 
spinors,
\begin{equation}\label{inhomDirWeil}
  \left(
    \frac{\partial}{\partial t}-\bm{\sigma}\bm{\nabla}
  \right)
  \eta_a+m_a \sigma_2 \eta_a^{*{}} =
  \mathrm{i} \phi_a.
\end{equation}
The solutions to Eq.~\eqref{inhomDirWeil} have the form (see, e.g., 
Ref.~\cite{FukYan03}),
\begin{align}
  \label{solWeileta}
  \eta_a(\mathbf{r},t)=\int\mathrm{d}^3\mathbf{r}'\mathrm{d}t'
  \mathcal{S}^\mathrm{ret}_a(\mathbf{r}-\mathbf{r}',t-t')
  \phi_a(\mathbf{r}',t'),
  \\
  \label{solWeileta*}
  \eta_a^{*{}}(\mathbf{r},t)=\int\mathrm{d}^3\mathbf{r}'\mathrm{d}t'
  \mathcal{R}^\mathrm{ret}_a(\mathbf{r}-\mathbf{r}',t-t')
  \phi_a(\mathbf{r}',t'),
\end{align}
where the retarded Green functions are expressed as (see also 
Ref.~\cite{FukYan03}),
\begin{align}\label{SRretM}
  \mathcal{S}_a^\mathrm{ret}(\mathbf{r},t)= &
  \mathrm{i} \tilde{\sigma}^\mu \partial_\mu D_a^\mathrm{ret}(\mathbf{r},t),
  \notag
  \\
  \mathcal{R}_a^\mathrm{ret}(\mathbf{r},t)= &
  \mathrm{i} m_a \sigma_2 D_a^\mathrm{ret}(\mathbf{r},t).
\end{align}
Here $\partial_\mu = (\partial_t,\bm{\nabla})$, $\tilde{\sigma}^\mu = 
(\widehat{\mathds{1}},\bm{\sigma})$, $\widehat{\mathds{1}}$ is the $2 \times 2$ 
unit matrix and $D_a^\mathrm{ret}(\mathbf{r},t)$ is given in 
Eq.~\eqref{retGfunScal}. One can check that Eq.~\eqref{solWeileta} 
and~\eqref{solWeileta*}, along with the definition of the retarded Green 
functions~\eqref{SRretM}, represent the solutions to Eq.~\eqref{inhomDirWeil} by 
means of direct substituting and using Eq.~\eqref{propD}.

Let us assume that the sources $\phi_a$ depend on time and spatial coordinates as 
in previous sections,
\begin{align}\label{Weilsour}
  \phi_a(\mathbf{r},t)= &
  \theta(t) \phi_a^{(0)} e^{-\mathrm{i} E t} \delta^3(\mathbf{r}),
  \notag
  \\
  \phi_a^{(0)} = & \sum_{\lambda=1}^{N}
  U^\dag_{a \lambda} f_\lambda \chi^{(0)}_\lambda,
\end{align}
where $\chi^{(0)}_\lambda$ is the time independent component of $\chi_\lambda$. In 
deriving of Eq.~\eqref{Weilsour} from Eq.~\eqref{phisour} we introduce the new 
quantities $f_\lambda \equiv f_\lambda^{(0)0}$, where $f_\lambda^{(0)0}$ is the 
time independent part of $f_\lambda^{0}$, to simplify the notations.

On the basis of Eqs.~\eqref{solWeileta}-\eqref{Weilsour} and using the technique of 
the previous sections we get the particles wave functions,
\begin{align}
  \label{solWeiletafin}
  \eta_a(\mathbf{r},t)= & \mathrm{i} \tilde{\sigma}^\mu \partial_\mu
  [e^{- \mathrm{i} E t + \mathrm{i} p_a r}]
  \frac{\phi_a^{(0)}}{4 \pi r},
  \\
  \label{solWeileta*fin}
  \eta_a^{*{}}(\mathbf{r},t)= & \mathrm{i} m_a \sigma_2
  e^{- \mathrm{i} E t + \mathrm{i} p_a r}
  \frac{\phi_a^{(0)}}{4 \pi r}.
\end{align}
To derive Eqs.~\eqref{solWeiletafin} and~\eqref{solWeileta*fin} we suppose that $t 
\gg r$. The derivatives in Eq.~\eqref{solWeiletafin} are applied on the exponent 
only because of the same reasons as in Sec.~\ref{DMT}.

Using Eqs.~\eqref{spinnupsirelM}, \eqref{solWeiletafin} and~\eqref{solWeileta*fin} 
as well as the following identity:
\begin{align}\label{ident}
  \left(
    \nu_\lambda^\mathrm{L}
  \right)^c = &
  \left(
    \nu_\lambda^c
  \right)^\mathrm{R}=
  \sum_{a=1}^{2N} U_{\lambda a}^{*{}}\psi_a^\mathrm{R},
  \notag
  \\
  \left(
    \nu_\lambda^\mathrm{L}
  \right)^c = & C
  \left(
    \overline{\nu_\lambda^\mathrm{L}}
  \right)^\mathrm{T},
\end{align}
we get the wave functions of $\nu_\lambda^\mathrm{L}$ and $\left( 
\nu_\lambda^\mathrm{L} \right)^c$ as
\begin{align}
  \label{nulambda}
  \nu_\lambda^\mathrm{L}(\mathbf{r},t) = &
  \frac{2E}{4 \pi r}
  e^{-\mathrm{i}E t}
  \sum_{a=1}^{2N}\sum_{\lambda'=1}^{N}
  e^{\mathrm{i} p_a r}
  U_{\lambda a} U_{a \lambda'}^\dag
  \notag
  \\
  & \times
  f_{\lambda'}
  \begin{pmatrix}
    0 \\
    \tilde{\chi}^{(0)}_{\lambda'}
  \end{pmatrix},
  \\
  \label{nulambdac}
  \left(
    \nu_\lambda^\mathrm{L}
  \right)^c(\mathbf{r},t) = &
  -\frac{2E}{4 \pi r}
  e^{-\mathrm{i}E t}
  \sum_{a=1}^{2N}\sum_{\lambda'=1}^{N}
  \frac{m_a}{2E} e^{\mathrm{i} p_a r}
  U_{\lambda a}^{*{}} U_{a \lambda'}^\dag
  \notag
  \\
  & \times
  f_{\lambda'}
  \begin{pmatrix}
    \chi^{(0)}_{\lambda'} \\
    0
  \end{pmatrix},
\end{align}
where $\tilde{\chi}^{(0)}_{\lambda'} = (1/2)[1-(\bm{\sigma}\mathbf{n})] 
\chi^{(0)}_{\lambda'}$. In deriving of Eq.~\eqref{nulambda} we suppose that
\begin{equation*}
  [E-p_a(\bm{\sigma}\mathbf{n})] \approx E [1-(\bm{\sigma}\mathbf{n})],
\end{equation*}
that is valid for relativistic neutrinos.

It is possible to construct two four-component Majorana spinors from two-component 
Weil spinors,
\begin{equation}\label{psi1psi2}
  \psi_a^{(1)} =
  \begin{pmatrix}
    \mathrm{i} \sigma_2 (\eta_a)^{*{}} \\
    \eta_a
  \end{pmatrix},
  \quad
  \psi_a^{(2)} =
  \begin{pmatrix}
    \mathrm{i} \sigma_2 \eta_a^{*{}} \\
    (\eta_a^{*{}})^{*{}}
  \end{pmatrix},
\end{equation}
where $\eta_a$ and $\eta_a^{*{}}$ are defined in Eqs.~\eqref{solWeiletafin} 
and~\eqref{solWeileta*fin}. One can see that these spinors satisfy the Majorana 
condition $\left[ \psi_a^{(1,2)} \right]^c = \psi_a^{(1,2)}$. We use the spinor 
$\psi_a^{(1)}$ to receive Eq.~\eqref{nulambda} and $\psi_a^{(2)}$ -- 
Eq.~\eqref{nulambdac}. In this case we get that $\nu_\lambda^\mathrm{L}$ and 
$\left( \nu_\lambda^\mathrm{L} \right)^c$ are obtained as a result of the evolution 
of particles emitted from the same source.

It should be mentioned that both $\nu_\lambda^\mathrm{L}$ and $\left( 
\nu_\lambda^\mathrm{L} \right)^c$ in Eqs.~\eqref{nulambda} and~\eqref{nulambdac} 
propagate forward in time. To explain this fact let us discuss the complex 
conjugated equation~\eqref{solWeileta}. Performing the same computations one 
arrives to the analog of Eq.~\eqref{nulambdac} which would depend on time as 
$e^{\mathrm{i}E t}$. However this wave function describes a particle emitted by the 
source different from that discussed here. Indeed, if we studied the complex 
conjugated Eq.~\eqref{solWeileta}, the integrand there would be 
$\mathcal{S}^\mathrm{ret*{}}_a \phi_a^{*{}}$. It would mean that the source in 
Eq.~\eqref{Weilsour} would be proportional to $\phi_a^{(0)*{}}e^{\mathrm{i} E t}$, 
that, in its turn, would signify that $\left( \nu_\lambda^\mathrm{L} \right)^c$ 
would be emitted in a process involving a lepton which is a charge conjugated 
counterpart to that discussed here.

Let us illustrate this problem on more physical example. Suppose  that we study a 
\emph{neutrino} emission in a process like the inverse $\beta$-decay,
\begin{equation}\label{exnuem}
  p + \ell_\lambda^{-{}} \to n + \nu_\lambda \Rightarrow
  n + \sum_{a=1}^{2N} U_{\lambda a}\psi_a^\mathrm{L},
\end{equation}
where $p$, $n$ and $\ell_\lambda^{-{}}$ stand for a proton, a neutron and for a 
negatively charged lepton. The complex conjugated Eq.~\eqref{solWeileta} would 
correspond to a process,
\begin{align}\label{exantinuem}
  n + \ell_\lambda^{+{}} \to p + \tilde{\nu}_\lambda \Rightarrow &
  p + \sum_{a=1}^{2N} U_{\lambda a}^{*{}}
  \left(
    \psi_a^\mathrm{L}
  \right)^c
  \notag
  \\
  \Rightarrow &
  p + \sum_{a=1}^{2N} U_{\lambda a}^{*{}}
  \psi_a^\mathrm{R},
\end{align}
where $\tilde{\nu}_\lambda$ and $\ell_\lambda^{+{}}$ denote an \emph{antineutrino} 
and a positively charged lepton, which is the charge conjugated counterpart to 
$\ell_\lambda^{-{}}$. In Eq.~\eqref{exantinuem} we use the facts that only 
left-handed interactions exist in nature and the fields $\psi_a$ describe Majorana 
particles. As one can see, Eqs.~\eqref{exnuem} and~\eqref{exantinuem} represent two 
different processes. Hence, if we used $(\eta_a)^{*{}}$ to obtain $\left( 
\nu_\lambda^\mathrm{L} \right)^c(\mathbf{r},t)$, i.e. after a beam of neutrinos 
passes some distance $r$, it would correspond to the initial 
reaction~\eqref{exantinuem} rather than~\eqref{exnuem}. Note that the same result 
also follows from Eq.~\eqref{psi1psi2} if we replace $\psi_a^{(1)} \leftrightarrow 
\psi_a^{(2)}$ there.

Let us suppose for simplicity that the momentum of the fermion $\ell_\lambda$ is 
parallel to the neutrino momentum. It takes place if a relativistic incoming lepton 
is studied. We also assume that this fermion is in a state with the definite 
helicity,
\begin{equation}\label{helstatefer}
  \frac{1}{2}[1-(\bm{\sigma}\mathbf{n})]\chi^{(0)}_\lambda=
  \chi^{(0)}_\lambda.
\end{equation}
This expression is again natural for the relativistic fermion $\ell_\lambda$. One 
can notice that interaction Lagrangian~\eqref{LagrintM} is written in terms of the 
left-handed chiral projections of $\ell_\lambda$. Therefore, if we study a 
relativistic lepton, it will have its spin directed oppositely to the particle 
momentum as one can see from Eq.~\eqref{helstatefer}.

Using Eqs.~\eqref{nulambda}, \eqref{nulambdac} and~\eqref{helstatefer} as well as 
the orthonormality of the two-component spinors $\chi^{(0)}_\lambda$, $\left( 
\chi^{(0)\dag}_\lambda \chi^{(0)}_{\lambda'} \right) = \delta_{\lambda\lambda'}$, 
we get the probabilities to detect $\nu_\lambda^\mathrm{L}$ and $\left( 
\nu_\lambda^\mathrm{L} \right)^c$ as
\begin{align}
  \label{Pnulambda}
  P_{\nu_\lambda^\mathrm{L}}(r) \sim &
  \sum_{ab=1}^{2N}\sum_{\lambda'=1}^{N}
  e^{\mathrm{i} (p_a-p_b) r}
  \notag
  \\
  & \times
  U_{\lambda a} U_{a \lambda'}^\dag U_{b \lambda'}^\dag U_{\lambda' b}
  |f_{\lambda'}|^2,
  \\
  \label{Pnulambdac}
  P_{\left( \nu_\lambda^\mathrm{L} \right)^c}(r) \sim &
  \sum_{ab=1}^{2N}\sum_{\lambda'=1}^{N}
  \frac{m_a m_b}{(2E)^2}
  e^{\mathrm{i} (p_a-p_b) r}
  \notag
  \\
  & \times
  U_{\lambda a}^{*{}} U_{a \lambda'}^\dag
  U_{b \lambda'}^\mathrm{T} U_{\lambda' b}
  |f_{\lambda'}|^2.
\end{align}
In Eqs.~\eqref{Pnulambda} and~\eqref{Pnulambdac} we drop the factor $(2E)^2/(4 \pi 
r)^2$.

We can see that Eqs.~\eqref{Pnulambda} and~\eqref{Pnulambdac} contain the 
oscillating exponent. This our result reproduces the usual formulae for neutrino 
oscillations in vacuum~\cite{Kob80}. It should be also noticed that the expressions 
for the probabilities depend on $r$ rather than on $t$ in contrast to our previous 
works~\cite{Dvo}. Note that the problem whether neutrino oscillations happen in 
space or in time was also discussed in Ref.~\cite{ShiNau07}. The similar coordinate 
dependence of the probabilities was obtained in 
Refs.~\cite{Kob82,GiuKimLeeLee93,GriSto96} where the problem of neutrino 
oscillations in vacuum was studied.

It was mentioned in Ref.~\cite{Kob82} that oscillations between active and sterile 
neutrinos are possible in case of the non-unitary matrix $(U_{\lambda a})$, i.e. 
the presence of only Majorana mass terms is not sufficient for the existence of 
this kind of transitions. The situation is analogous to that considered in the 
pioneering work~\cite{PontecorvoOsc} where oscillations between neutrinos and 
antineutrinos were studied. In Eq.~\eqref{Pnulambdac} we obtain that the 
probability to detect $\left( \nu_\lambda^\mathrm{L} \right)^c$ is suppressed by 
the factor $m_a m_b/E^2$. It was also mentioned in Refs.~\cite{Kob82,nutoantinu}.

\section{Discussion and Conclusions\label{CONCLUSION}}

In conclusion we mention that we studied the evolution of the systems of mixed 
particles emitted by classical sources. The cases of both scalar and spinor 
particles were considered. The fields distributions of particles were found for the 
arbitrary sources dependences on time and spatial coordinates.
Note that the expressions for the fields distributions were received directly from 
the Lorentz invariant wave equations and thus they are valid for arbitrary energies 
of emitted particles. It should be also mentioned that all the calculations were 
carried out in $(3+1)$ dimensions.

We demonstrated that the expressions for energy densities, in the scalar fields 
case, and intensities, in the spinor fields case, coincide with the usual 
transition and survival probabilities of neutrino flavor oscillations in vacuum 
derived within the quantum mechanical approach. Since we study the process of 
neutrino oscillations on the basis of the field theory method we reproduce not only 
the quantum mechanical probabilities as the leading terms, but also one can derive 
the corrections to these expressions. It is known (see Refs.~\cite{BlaVit95,Dvo}) 
that the Pontecorvo formula for neutrino oscillations has the corrections 
$\sim(m_a/E)^2$, that are small for ultrarelativistic neutrinos. In the present 
work we showed that the analogous terms also appear in the expressions for 
probabilities [see, e.g., Eqs.~\eqref{phi12endens} and~\eqref{spinint1}]. In 
addition, in Sec.~\ref{GMT} we considered the possibility of the appearance of 
antiparticles in the initial particles beam.

In our previous works~\cite{Dvo} we studied neutrino flavor oscillations in vacuum 
and in various external fields within the field theoretical approach. In those 
papers we solved the initial condition problem for the wave equations describing 
the evolution of neutrino \emph{mass} eigenstates. To study the dynamics of the 
particles one should use Pauli-Jordan function as the Green function for the 
considered initial condition problem. As a result we received the final wave 
functions corresponding to the time $t$ of the system evolution. It was shown that 
the corresponding probabilities to detect a \emph{flavor} neutrino coincide with 
usual probabilities of neutrino oscillations derived in the Schr\"odinger equation 
approach. Therefore we obtained that neutrino oscillations happen in time. The 
weakness of that method was that to get the stable oscillations picture, i.e. 
reproduce the Pontecorvo formula, one should prepare very broad initial wave 
packet. Physically it would correspond to a permanent, or rather long-continued, 
neutrino source.

One can also treat neutrino flavor oscillations in vacuum describing the 
propagation of neutrino \emph{mass} eigenstates as the internal lines in a Feynman 
diagram~\cite{Kob82,GiuKimLeeLee93,GriSto96}. Thus a Feynman propagator serves as a 
Green function in that approach. In that method one replaces \emph{flavor} 
eigenstates with \emph{mass} eigenstates at the neutrino emission and detection 
points. It is possible to say that one describes neutrino oscillations without 
resort to flavor neutrinos. It was found that transition and survival probabilities 
depend on the distance between emission and detection points. Although that 
approach seems to be quite natural it encounters some problems. For example, it is 
unclear (i) if different \emph{mass} eigenstates should have equal energies, (ii) 
if one loses coherence when the energy is measured with high accuracy, (iii) if the 
usual Pontecorvo formula is unique because the non-standard oscillations formulae 
are possible in that approach (see the criticism in Ref.~\cite{Beu03}).

In the present work we developed another approach to the neutrino flavor 
oscillations problem. In frames of our method we derived the coordinate dependent 
transition and survival probabilities. We examined the relaxation phenomena 
occurring when a neutrino signal arrives to a detector. It was revealed that the 
stable oscillations picture could be implemented if the time after the neutrino 
emission is much greater than the distance between a source and a detector. It 
means that the quantum mechanical transition probability formula is valid for a 
rather long-continued neutrino source. This result, in a sense, confirms the 
results of our previous works~\cite{Dvo} (see also above). Note
that one can come the similar conclusions basing on
Ref.~\cite{IoaPil99} where neutrino oscillations were studied on
the basis of an exactly solvable model.

The weakness of the present method is that in various processes
where a neutrino participates, particles belonging to different
flavors can be emitted (see, e.g., Ref.~\cite{Shr80}). For
instance, there are the following decay channels of a charged
pion: $\pi^{+{}} \to \mu^{+{}} + \nu_{\mu}$ and $\pi^{+{}} \to
e^{+{}} + \nu_{e}$. Therefore in frames of our model the classical
sources of both muon and electron neutrinos would correspond to
this process. However the latter channel is suppressed by the
factor $10^4$~\cite{Yao06}. Thus we can suppose that only one type
of neutrino sources is presented. Nevertheless one can formally
consider the sources of all neutrino species in frames our method.
The amplitudes $f_\lambda^{(0)\mu}$ of the sources should be taken
either from the quantum field theory of the neutrino emission or
from an experiment.

Note that neutrino flavor oscillations in the model with a 
source and a detector were described in Ref.~\cite{KieWei98}. In that paper the 
source of particles was also assumed to be spatially localized.
However its time dependence was chosen to be gaussian that is
different from our approach.

Summarizing we can say that our method for the description of the
flavor neutrinos evolution can be useful when we have spatially
localized neutrino sources which start to emit neutrinos at some
moment of time. One of the possible examples is the neutrinos
emission in a supernova explosion.

\begin{acknowledgments}
The work has been supported by the Academy of Finland under the
contract No.~108875 and by the Conicyt (Chile), Programa
Bicentenario PSD-91-2006. The author is thankful to the Russian
Science Support Foundation for a grant as well as to Jukka
Maalampi (University of Jyv\"{a}skyl\"{a}) and Walter Grimus
(University of Vienna) for helpful discussions.
\end{acknowledgments}

\appendix

\section{Evaluation of integrals\label{ERROR}}

It is interesting to evaluate the inexactitude which is made when we approach to 
the limit $x_m\to\infty$ in Eq.~\eqref{int2}. Let us discuss two functions
\begin{align}
  \label{errorF}
  F(r,t)= & \int_0^{x_m}\mathrm{d}x
  \frac{J_1(m x)}{\sqrt{r^2+x^2}}e^{\mathrm{i}E\sqrt{r^2+x^2}}
  \notag
  \\
  = &
  \int_0^{y_m}\mathrm{d}y
  \frac{J_1(\rho y)}{\sqrt{1+y^2}}e^{\mathrm{i}\mathcal{E}\sqrt{1+y^2}},
  \\
  \label{errorF0}
  F_0(r)= & \frac{1}{\rho}(e^{\mathrm{i}\mathcal{E}}-e^{\mathrm{i}P}).
\end{align}
These functions are proportional to $\mathcal{I}_2$ in Eqs.~\eqref{int2} 
and~\eqref{int2fin} respectively. In Eqs.~\eqref{errorF} and~\eqref{errorF0} we use 
dimensionless parameters $\rho=m r$, $\mathcal{E}=E r=\gamma\rho$, $\gamma=E/m$, 
$y_m=\sqrt{(t/r)^2-1}$ and $P=\sqrt{\mathcal{E}^2-\rho^2}$.

In case we study neutrinos, we get that $\mathcal{E} \gg \rho \gg 1$ in almost all 
realistic situations. For example, suppose we study a neutrino emitted in a 
supernova explosion is our Galaxy. The typical distance is $r \sim 
10\thinspace\text{kpc}$. Taking $m \sim 1\thinspace\text{eV}$ and $E \sim 
10\thinspace\text{MeV}$, we receive that $\mathcal{E} \sim 10^{35}$ and $\rho \sim 
10^{28}$.

Basing on the analysis of Sec.~\ref{SCALAR} we can rewrite Eq.~\eqref{errorF} as
\begin{align}\label{deltaF}
  F(r,t)= & F_0(r)-\delta F,
  \notag
  \\
  \delta F= & \int_{y_m}^{+\infty}\mathrm{d}y
  \frac{J_1(\rho y)}{\sqrt{1+y^2}}e^{\mathrm{i}\mathcal{E}\sqrt{1+y^2}}.
\end{align}
Using the fact that $\rho \gg 1$, $y_m \gg 1$ and the representation for the Bessel 
function,
\begin{equation}
  J_1(z) \approx
  \sqrt{\frac{2}{\pi z}}\cos
  \left(
    z-\frac{3\pi}{4}
  \right)
  \quad
  \text{at}
  \quad
  z \to +\infty,
\end{equation}
we obtain for the function $\delta F$ in Eq.~\eqref{deltaF} the following 
expression:
\begin{align}
  \delta F \approx & -\frac{1}{\sqrt{2\pi\rho}}
  \big\{
    \mathrm{ci}([\gamma+1]\rho y_m)+\mathrm{ci}([\gamma-1]\rho y_m)
    \notag
    \\
    & +
    \mathrm{i}
    [\mathrm{si}([\gamma+1]\rho y_m)+\mathrm{si}([\gamma-1]\rho y_m)]
  \big\},
\end{align}
where $\mathrm{ci}(z)$ and $\mathrm{si}(z)$ are cosine and sine integrals. Using 
the asymptotic expression,
\begin{equation}
  |\mathrm{ci}(z)| \sim |\mathrm{si}(z)| \sim \frac{1}{z}
  \quad
  \text{at}
  \quad
  z \to +\infty,
\end{equation}
we obtain that the function $\delta F$ approaches to zero as $1/(y_m \rho^{3/2})$ 
at great values of $y_m$ and $\rho$. Note that this result remains valid for a 
particle with an arbitrary $\gamma$ factor, i.e. rapid oscillations of the function 
$\delta F$ will attenuate even for slow particles. This analysis substantiates the 
approximations made in Sec.~\ref{SCALAR}.

\begin{figure}
  \centering
  \includegraphics[scale=.45]{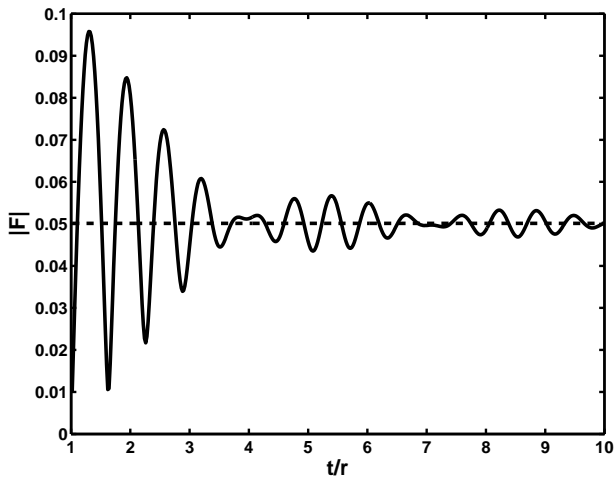}
  \caption{The absolute values of the functions $F(r,t)$ and $F_0(r)$
  versus $t$.}\label{error}
\end{figure}

Finally let us illustrate the behavior of the functions $F(r,t)$ and $F_0(r)$. On 
Fig.~\ref{error}
we present the absolute values of these functions versus $t$. This figure is 
plotted for $\rho=1$ and $\mathcal{E}=10$. The solid line is the absolute value of 
the function $F(r,t)$ and the dashed line -- $F_0(r)$. As we mention in 
Sec.~\ref{SCALAR} the relaxation phenomena occur when $t\gtrsim r$. It can be also 
seen on Fig.~\ref{error}. It is possible to notice that $|F(r,t)|\to|F_0(r)|$ at 
great values of $t$ as it is predicted in Sec.~\ref{SCALAR}.

\end{document}